\documentclass{elsart}
\usepackage{changebar,float,epsfig}

\begin{document}
\begin{frontmatter}

\title{Nonlinear analysis of bivariate data with cross recurrence plots}
\author{Norbert Marwan\corauthref{cor}},
\corauth[cor]{Corresponding author.}
\author{J\"urgen Kurths}
\address{Department of Physics, University of Potsdam, Potsdam 14415, Germany}

\begin{abstract}
We use the extension of the method of recurrence plots to 
cross recurrence plots (CRP) 
which enables a nonlinear analysis of bivariate data. To quantify CRPs, 
we develop further three measures of complexity mainly basing on diagonal 
structures in CRPs. The CRP analysis of prototypical model systems 
with nonlinear interactions demonstrates that this technique enables 
to find these nonlinear interrelations from bivariate time series, whereas 
linear correlation tests do not. Applying the CRP analysis to climatological 
data, we find a complex relationship between rainfall and El Ni\~no data.
\end{abstract}
\end{frontmatter}

\section{Introduction}

A major task in bi- or multivariate data analysis is to compare or to 
find interrelations 
in different time series. Often, these data are gained from
natural systems, which show generally nonstationary and
complex behaviour. Furthermore, these systems are often observed by very 
few measurements providing short data series. Linear approaches of 
time series analysis are often not sufficient to analyse this kind of 
data. In the last two decades a great variety
of nonlinear techniques has been developed to analyse 
data of complex systems (cf.~\cite{abarbanel93,kantz97}). 
Most popular are methods to estimate fractal dimensions, 
Lyapunov exponents or mutual information 
\cite{kantz97,kurths87,mandelbrot82,wolf85}. 
However, most of these methods need long data series. 
The uncritical application of these methods especially to
natural data often leads to pitfalls.

To overcome the difficulties with nonstationary and rather short 
data series, the method of {\it recurrence plots (RP)}
has been introduced \cite{casdagli97,eckmann87,koebbe92}.
An additional quantitative analysis of recurrence plots
has been developed to detect transitions (e.\,g.~bifurcation points) 
in complex systems \cite{trulla96,webber94,zbilut92,marwan2002herz}. 
An extension of the method of recurrence plots to 
cross recurrence plots enables to investigate 
the time dependent behaviour of two
processes which are both recorded in a single time series 
\cite{zbilut98,marwan2002npg}.
The basic idea of this approach is to compare the phase space
trajectories of two processes in the same phase space. 
The aim of this work is to develop further new measures of complexity,
which are based on cross recurrence plots and to evaluate
the similarity of the considered systems. This 
nonlinear approach enables to identify epochs where there are
linear and even nonlinear interrelations between both systems.

Firstly, we give an overview about recurrence plots and
cross recurrence plots and, than, we develop further new measures 
of complexity. Lastly, we apply the method to two model 
systems and to natural data.

\section{Recurrence Plot}

The recurrence plot (RP) is a tool in order to visualize
the dynamics of phase space trajectories and was 
firstly introduced by Eckmann et al.~\cite{eckmann87}.
Following Takens' embedding theorem \cite{takens81}, 
the dynamics can be appropriately presented by a reconstruction
of the phase space trajectory $\vec x(t)$ from a time 
series $u_k$ (with a sampling time $\Delta t$) by using an embedding 
dimension $m$ and a time delay $\tau$ 
\begin{equation}\label{embedding}
\vec x(t)=\vec x_i=\left( u_i, u_{i+\tau}, \dots,  
u_{i+(m-1)\,\tau} \right), \quad t=i \, \Delta t.
\end{equation}
The choice of $m$ and $\tau$ are based
on standard methods for detecting these parameters like
method of false nearest neighbours (for $m$) and
mutual information (for $\tau$), which ensures the entire
covering of all free parameters and avoiding of autocorrelation
effects \cite{kantz97}. 

The recurrence plot is defined as
\begin{equation}
\mathbf{R}_{i,\,j}=\Theta\left(\varepsilon_i-\|\vec x_i-\vec x_j\|\right), 
\end{equation}
where $\varepsilon_i$ is a predefined 
cut-off distance, $\| \cdot \|$ is the norm (e.\,g.~the Euclidean norm) and
$\Theta(x)$ is the Heaviside function. The values
{\it one} and {\it zero} in this matrix can be simply 
visualized by the colours black and white. 
Depending on the kind of the application, $\varepsilon_i$ can be a fixed
value or it can be changed for each $i$
in such a way that in the ball with the radius $\varepsilon_i$
a predefined amount of neighbours occurs. The latter will provide a constant 
density of recurrence points in each column of the RP.
Such a RP exhibits characteristic large-scale and
small-scale patterns which are caused by typical dynamical
behavior \cite{eckmann87,marwan2002herz,webber94}, e.\,g.~diagonals (similar
local evolution of different parts of the trajectory) or
horizontal and vertical black lines (state does not change for
some time).
A single recurrence point, however, contains no information about the
state itself.

As a quantitative extension of the method of recurrence plots, 
the {\it recurrence quantification analysis (RQA)} was introduced 
by Zbilut and Webber \cite{webber94,zbilut92}. 
This technique defines several measures mostly based on diagonal oriented lines in
the recurrence plot: {\it recurrence rate}, {\it determinism},
{\it maximal length of diagonal structures}, {\it entropy} and
{\it trend}. The {\it recurrence rate} is the ratio of all recurrent states 
(recurrence points) to
all possible states and is therefore the probability of the recurrence of
a certain state. Stochastic behaviour causes very short diagonals,
whereas deterministic behaviour causes longer diagonals. Therefore, the
ratio of recurrence points forming diagonal structures to all recurrence
points is called the {\it determinism} (although this measure does not
really reflect the determinism of the system). 
Diagonal structures show the range in which a piece of the trajectory 
is rather close to another one at different time. The
{\it diagonal length} is the time span they will be close to each other and 
their mean can be interpreted as the mean prediction time. The inverse
of the maximal line length can be interpreted to be directly
related with the maximal positive Lyapunov exponent 
\cite{eckmann87,trulla96,choi99}; in this interpretation it is
assumed that the considered system is chaotic and has no stochastic 
influences.
Since real (natural) systems are always affected by noise, we suggest that
this measure has to be interpreted in a more statistical way, 
for instance as a prediction time. However, if we consider a
chaotic system, the maximal positive Lyapunov exponent is much 
more reflected in the distribution of the line lengths.
The {\it entropy} is defined as the Shannon entropy in the histogram 
of diagonal line lengths. Stationary systems will deliver rather homogeneous recurrence 
plots, whereas nonstationary systems cause changes in the distribution
of recurrence points in the plot visible by brightened areas.
For example, a simple drift in the data causes a paling of the recurrence plot
away from the main diagonal to the edges. The parameter 
{\it trend} measures this effect by diagonal wise computation of 
the diagonal recurrence density and its
linear relation to the time distance of these diagonals to the main diagonal.

\section{Cross Recurrence Plot}

Analogous to Zbilut et al.~\cite{zbilut98}, we will use the recently 
expanded method of recurrence plots to the method of 
{\it cross recurrence plots}, which compares the dynamics 
represented in two time series. Herein, both time series are 
simultaneously embedded in the same phase space. The test
for closeness of each point of the first trajectory $\vec x_i$ ($i=1 \dots N$) 
with each point of the second trajectory $\vec y_j$ ($j=1 \dots M$)
results in a $N \times M$ array
$\mathbf{CR}_{i,\,j}=\Theta(\varepsilon-\|\vec x_i-\vec y_j\|)$ called the 
cross recurrence plot (CRP). Visual inspection of CRPs already 
reveals valuable information about the relationship between both  
systems. Long diagonal structures show similar
phase space behaviour of both time series. It is obvious, that if
the difference of both systems vanishes, the main diagonal
line will occur black. An additional time dilatation or compression
of one of these similar trajectories causes a distortion of
this diagonal line \cite{marwan2002npg}. In the following, we suppose
that both systems do not have differences in the time scale and
have the same length $N$, hence, the CRP is a $N \times N$ array and
an increasing similarity between both systems causes a raising of 
the recurrence point density along the main diagonal until a black 
straight main diagonal line occurs (cf.~Fig.~\ref{CRQAsin}). 
Finally, the CRP compares the considered systems and allows 
us to benchmark their similarity.

\section{Complexity measures based on cross recurrence plots}

Next, we will define some modified RQA measures for quantifying the similarity
between the phase space trajectories. Since we use the occurrence of the
more or less discontinuous main diagonal as a measure for similarity,
the modified RQA measures will be determined for each diagonal 
line parallel to the main diagonal, hence, as functions
of the distance from the main diagonal. Therefore, it is also possible to 
assess the similarity in the dynamics depending on a certain delay. 

We analyze the distributions of the diagonal line 
lengths $P_{t}(l)$ for each diagonal parallel to the main diagonal.
The index $t \in [ -T \ldots T]$ marks the number of the diagonal line,
where $t=0$ marks the main diagonal, $t>0$ the diagonals above and
$t<0$ the diagonals below the main diagonal, which represent
positive and negative time delays, respectively.

The recurrence rate $RR$ is now defined as
\begin{equation}
RR(t) = \frac{1}{N-t} \sum_{l=1}^{N-t} l\, P_{t}(l)
\end{equation}
and reveals the probability of occurrence of 
similar states in both systems with a given delay $t$.
A high density of recurrence points in a diagonal results in
a high value of $RR$. This is the case for systems whose
trajectories often visit the same phase space regions.

Analogous to the RQA the determinism 
\begin{equation}
DET(t) = \frac{\sum_{l=l_{min}}^{N-t} l\, P_{t}(l)}
{\sum_{l=1}^{N-t} l\, P_{t}(l)}
\end{equation}
is the proportion of recurrence points forming 
long diagonal structures of all recurrence 
points. Stochastic as well as heavily fluctuating data cause none or only
short diagonals, whereas deterministic systems cause longer 
diagonals. If both deterministic systems have the same or
similar phase space behaviour, i.\,e.~parts of the phase space trajectories
meet the same phase space regions during certain times, the
amount of longer diagonals increases and the amount of smaller 
diagonals decreases.

The average diagonal line length 
\begin{equation}
L(t) = \frac{\sum_{l=l_{min}}^{N-t} l\, P_{t}(l)}
{\sum_{l=l_{min}}^{N-t} P_{t}(l)}
\end{equation}
reports the duration of
such a similarity in the dynamics. A high coincidence of 
both systems increases the length of these diagonals. 

High values of $RR$ represent high probabilities of the occurrence
of the same state in both systems, high values of $DET$ and $L$ represent 
a long time span of the occurrence of a similar dynamics in both systems.
Whereas $DET$ and $L$ are sensitive to fastly and highly fluctuating data,
$RR$ measures the probabilities of the occurrence 
of the same states in spite of these high fluctuations (noisy data).
It is important to emphasize
that these parameters are statistical measures and that their validity 
increases with the size of the CRP.

Compared to the other methods, this CRP technique has important advantages.
Since all parameters are computed
for various time delays, lags can be identified and causal links
proposed. An additional analysis with opposite signed second time series
allows us to distinguish positive and negative relations.
To recognize the measures for both cases,
we add the index $+$ to the measures 
for the positive linkage and the index $-$ for
the negative linkage, e.\,g.~$RR_+$ and $RR_-$.
A further substantial advantage of our method is the capability 
to find also nonlinear similarities in short and nonstationary 
time series with high noise levels as they typically occur, e.\,g., in
biology or earth sciences. However, the shortness and nonstationarity 
of data limits this method as well. One way to reduce problems that
occur with nonstationary data is the alternative choice of the neighbourhood
as a fixed amount of neighbours in the ball with a 
varying radius $\varepsilon$. A further major aspect is the
reliability of the found results. Until a mature statistical test
is developed, a first approach could be a surrogate test.

In the next section we apply these measures of complexity to
prototypical model systems and to real data.

\section{Examples illustrating the CRP}

\subsection{Noisy periodic data}

\begin{figure}[tbph]
\centering \epsfig{file=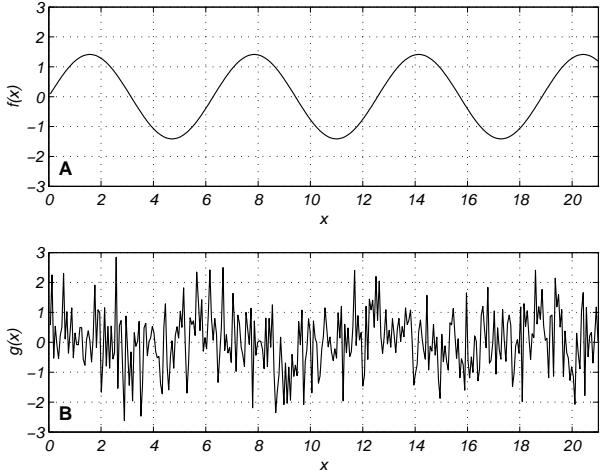, width=8cm}
\caption{Two delayed sine functions, one
of them corrupted by additive white noise (B).}\label{DATAsin}
\end{figure}

First, we consider a classical example to check whether our
technique is there compatible with linear statistical tools:
two sine functions $f(x)$ and $g(x)$ with the same period ($2\,\pi$), 
whereby the second function $g(x)$ is shifted by $\pi/2$ and strongly
corrupted by additive Gaussian white noise $\xi \in [-1, 1]$; 
the signal to noise ratio is 0.5 (Fig.~\ref{DATAsin}). 
Both time series have a length of 500 data points with a sampling rate
of $2\pi/100$.

\begin{figure}[tbph]
\centering \epsfig{file=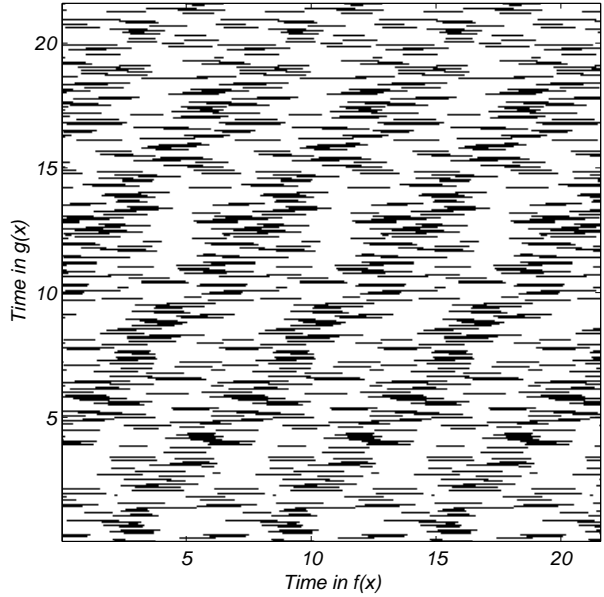,width=8cm}
\caption{Cross recurrence plot for two delayed sine functions
(Fig.~\ref{DATAsin})
with an embedding of $m=3$, $\tau=\pi/2$ and $\varepsilon=1.5$. 
The diagonal lines in the 
CRP result from similar phase space behaviour of 
both functions.}\label{CRPsin}
\end{figure}

We apply our analysis with $m=3$, $\tau=\pi/2$ and $\varepsilon=1.5$
(fixed radius, Euclidean distance). 
The CRP shows diagonal structures separated by gaps (Figs.~\ref{CRPsin}). 
These gaps are the result of the high fluctuation of the noisy
sine function. Due to the periodicity of these functions, 
the diagonals have a constant distance
to each other equal to the value of the period $\lambda=2\,\pi$.
The interrupted diagonal structures consist of a number of
short diagonals. However, these are long enough to achieve
significant maxima in the measures $RR$, $DET$ and $L$.

\begin{figure}[tbph]
\centering \epsfig{file=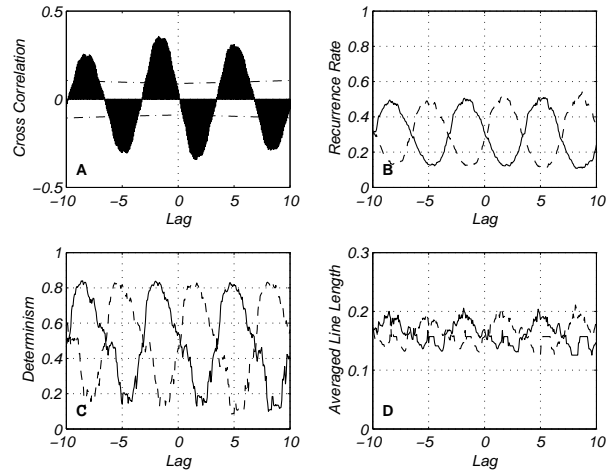,width=8cm}
\caption{Cross-correlation (A), $RR$ (B), $DET$ (C) and $L$ (D) for
two delayed sine functions. $L$ has the unit of time. 
The solid black lines show positive relation, the
dashed lines show negative relation. The dash-dotted line in (A) 
marks the 5\% confidence interval.
All functions (A)--(D) detect the correlation after a lag of $\pi/2$.}\label{CRQAsin}
\end{figure}

As expected, in this example the classical cross-correlation function 
shows a significant correlation
after a lag of $\pi/2$ (Fig.~\ref{CRQAsin}A). 
The $RR$, $DET$
and $L$ functions also show maxima for positive and negative relation
between $f(x)$ and $g(x)$. These maxima occur with the same lags 
$\pi/2$ like the linear 
correlation test (Fig.~\ref{CRQAsin}B-D). Despite the high
noise level, these measures find the correlation. Hence, the 
result of this CRP analysis agrees with the linear correlation analysis.

Due to the noisy data, the trajectories strongly fluctuate
in the phase space. Therefore, only short diagonal lines in the CRP occur
and the means of the measures $DET$ and $L$ have (relative) small values.

\subsection{System with nonlinear correlations}

\begin{figure}[tbph]
\centering \epsfig{file=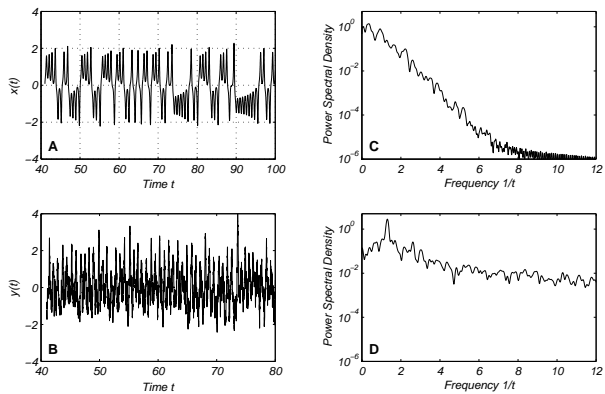, width=8cm}
\caption{(B) Time series of a nonlinear
related system consisting of a driven first order autoregressive process,
forced by the squared (A) $x$-component of the Lorenz system ($\kappa=0.2$).
The major periods (frequencies) are 2.9 (0.34) and 1.1 (0.94) for $x$ (C) 
and 0.77 (1.30) and 0.96 (1.05) for $y$ (D).}\label{DATAar}
\end{figure}

The next example is concerned to a nonlinear interrelation
between systems. We will study this interrelation by using
a standard linear method (cross correlation), a standard 
method from nonlinear data analysis (mutual information, 
cf.~\cite{kantz97}) and the new proposed measures.
We consider linear correlated noise (autoregressive process),
which is nonlinearly coupled with the $x$-component
of the Lorenz system $x(t)$ (solved with an ODE solver for the standard 
parameters $\sigma=10$, $r=28$, $b=8/3$ and a time resolution of 
$\Delta t = 0.01$, \cite{lorenz63,argyris94}). 
We use a first order autoregressive process $y_n$ and force it
with the squared $x$-component
\begin{equation}\label{eq_ar}
y_n=0.86 \, y_{n-1} + 0.500 \, \xi_n + \kappa \, x_n^2,
\end{equation}
where $\xi$ is Gaussian white noise and $x_n$ ($x(t) \rightarrow x_n$, $t=n \Delta t$)
is normalized to standard deviation $\sigma=1$ (Fig.~\ref{DATAar}).
The data length is 8,000 points. The coupling $\kappa$
is realized without any lag. In order to study the behaviour of the 
proposed measures as a function of the coupling strength, we compute
the CRPs for $\kappa \in [0, 3]$ and for 500 independent realizations.
The major periods of the system $x$ are 2.9 and 1.1, whereas the 
major periods of the selected realization of the system $y$
shown in Fig.~\ref{DATAar} are 0.77, 0.96 and 0.59 
(ordered from highest to lower).

\begin{figure}[tbph]
\centering \epsfig{file=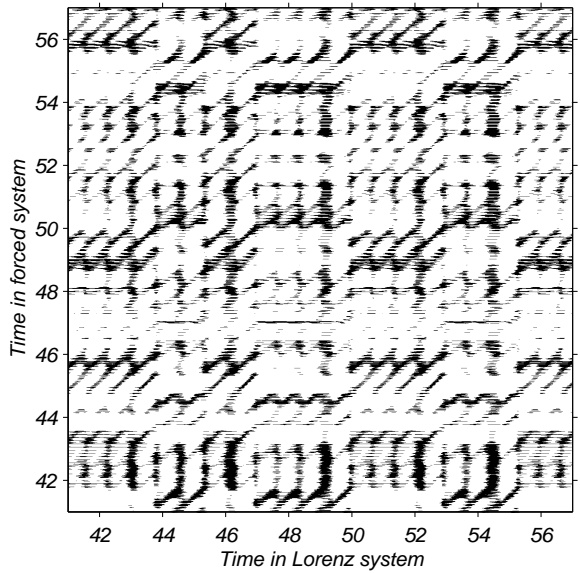, width=8cm}
\caption{Cross recurrence plot for
the forced autoregressive process $y$ (Fig.~\ref{DATAar}B) 
and the forcing function ($x$-component of the Lorenz system, 
Fig.~\ref{DATAar}A) for a coupling strength $\kappa=0.2$
and an embedding $m=5$, $\tau=10$, $\varepsilon=2$.}\label{CRPar}
\end{figure}

\begin{figure}[tbph]
\centering \epsfig{file=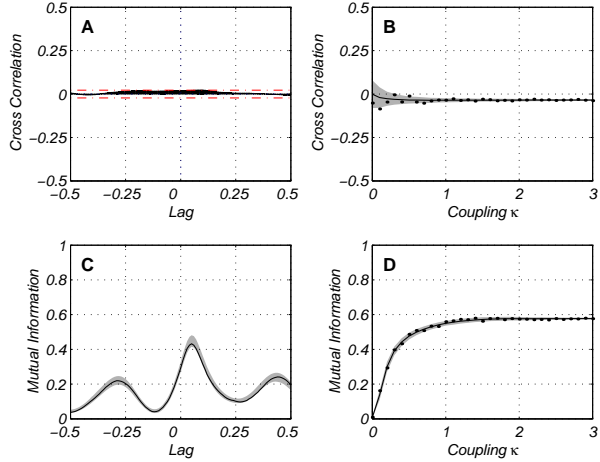, width=8cm}
\caption{Cross-correlation (A, B) and mutual information (C, D) for
the forced autoregressive process and the forcing function; A and C
represents the measures for one realization as functions of the 
delay and for a coupling $\kappa=0.2$, B and D represents the
measures for one realization (dots) and averaged (line) 
as functions of the coupling strength $\kappa$ (for a delay of zero). 
The dash-dotted lines in A mark the significance level of 5\,\%
for the linear correlation between $x$ and $y$, the gray bands in B, C and D mark 
the $2\sigma$ margin of the distributions of the measures 
gained from the 500 realizations.
The cross-correlation function does not find a significant 
correlation, but the mutual information shows significant
interrelations between $x$ and $y$ at delays of $0.05$, 
$0.4$ and $-0.3$. The correlation coefficient does
not clearly change for a growing coupling strength (B),
however, the mutual information monotonically increases
with a growing coupling strength $\kappa$ up to $\kappa=1$ and
does not change for $\kappa > 1$ (D). }\label{CRQAar}
\end{figure}

The cross correlation analysis of $x$ and $y$ do not reveal 
a significant linear correlation between them (Fig.\ref{CRQAar} A, B).
The linear correlation does not increase for a growing  coupling
strength $\kappa$. However, the mutual information 
shows a strong dependence between $x$ and $y$ at delays of $0.05$, $-0.29$ and 
$0.44$ (Fig.\ref{CRQAar} C, D). 
This measure increases for a growing coupling. Analogous results
can also be found with other nonlinear techniques 
which are designed for the study of 
interrelations as described in \cite{schreiber2000,schmitz2000}.

The CRP of the driven AR-process (Eq.~\ref{eq_ar}) with the $x$-component 
of the Lorenz system ($m=5$, $\tau=10$, $\varepsilon=2$)
contains a lot of longer diagonal lines, which represent time
ranges in which both systems have a similar phase space dynamics 
(Fig.~\ref{CRPar}). The results of the quantitative analysis of the CRP is
strongly different from those of the linear analysis. It is important
to note that the linear correlation analysis is here not able to  
detect any significant coupling or correlation 
between both systems (Fig.~\ref{CRQAar}A and B). 

\begin{figure}[tbph]
\centering \epsfig{file=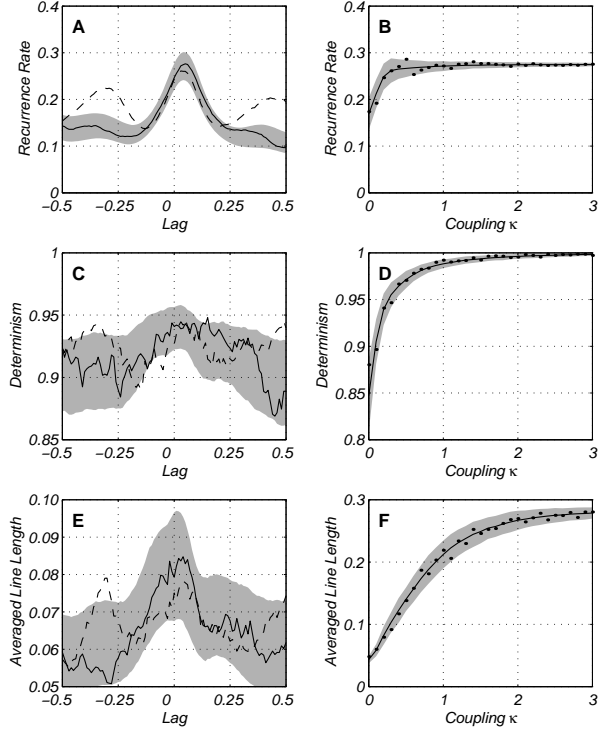, width=8cm}
\caption{$RR$ (A, B), $DET$ (C, D) and $L$ (E, F) for
the forced autoregressive process and the forcing function 
($L$ has the unit of time). The solid lines show positive relation,
the dashed lines show negative relation. 
The gray bands mark the $2\sigma$ margin of the distributions
of the measures gained from the 500 realizations; only
the $2\sigma$ margins for $RR_+$, $DET_+$ and $L_+$ are shown.
$RR_+$/ $L_+$ and $RR_-$/ $L_-$ have clear maxima 
for a lag about $0.05$,  $RR_-$ and $L_-$ have additionally
maxima at $0.4$ and $-0.3$, which is the similar behaviour as the
mutual information. The dependence from the coupling strength $\kappa$
is slightly different. Whereas $RR$ increases rather fast with 
growing $\kappa$ (B), $DET$ increases slower (D)  and $L$ increases
much slower (F) with growing $\kappa$.
Since the maxima occur for $RR_+$, $DET_+$ and $L_+-$ as well as
for $RR_-$, $DET_-$ and $L_-$, the found relation is of the kind of 
an even function.}\label{CRQAar2}
\end{figure}

Our measures of complexity exhibit the following:
$RR$ and $L$ exhibit
maxima at a lag of about $0.05$ for $RR_+$/ $L_+$ and $RR_-$/ $L_-$ 
and additionally at $0.45$ and $-0.32$ for $RR_-$/ $L_-$ (Figs.~\ref{CRQAar2}A, E). 
The delay of about $0.05$ stems from the auto correlation of $y$ and 
approximately corresponds to its correlation time $\Delta t/\ln{0.86}=0.066$.
The other both delays are in the sum $0.77$ which suggests, that they
are due to an interference of the main periods of the systems.
$DET_+$ and $DET_-$ has also maxima at these delays, but these
maxima are not significant in the sense that the values exceed the
$2\sigma$-level of the $DET$ distribution gained from 500 realizations
(Figs.~\ref{CRQAar2}C). This is due to the 
rapid fluctuating of $y$ and, thus, the less amount of longer 
diagonal structures ($l>3$). The reconstructed phase-space
trajectories of $x$ and $y$ do not run parallel for some time.

The three measures have a slightly different dependence on the 
coupling strength $\kappa$: whereas $RR$ increases rather fast with 
growing $\kappa$, $DET$ increases slower and $L$ increases
much slower with growing $\kappa$ (Figs.~\ref{CRQAar2}B, D, F). 
In comparison with the mutual information, the proposed measures
have a similar regime, but especially $DET$ and $L$, spread stronger. 
However, this spread depends on the length of the considered data and
decreases for longer data sets. 

Finally we can infer, that the measures $RR$ and $L$ are suitable
in order to find the nonlinear relation between the considered data 
series, where the linear analysis is not able to detect this relation. 
In this example, $DET$ does not reveal the nonlinear relation, because
the rapidly fluctuation in $y$ kicks away the reconstructed phase-space
trajectory from the parallel running to the trajectory of $x$. Since the 
result is rather independent of the sign of the second data before the
embedding, the found relation is of the kind of an even function.

\subsection{Climatological data}\label{sec_soi}

The last example shows the potential of the CRPs in order
to find interrelations in natural data. We investigate, 
whether there is a relation between the precipitation 
in an Argentinian city and the El Ni\~no/ Southern Oscillation
(ENSO). Power spectra analysis of local rainfall data
found periodicities of 2.3 and 3.6 years within 
the ENSO frequency band \cite{trauth2000}.

\begin{figure}[tbph]
\centering \epsfig{file=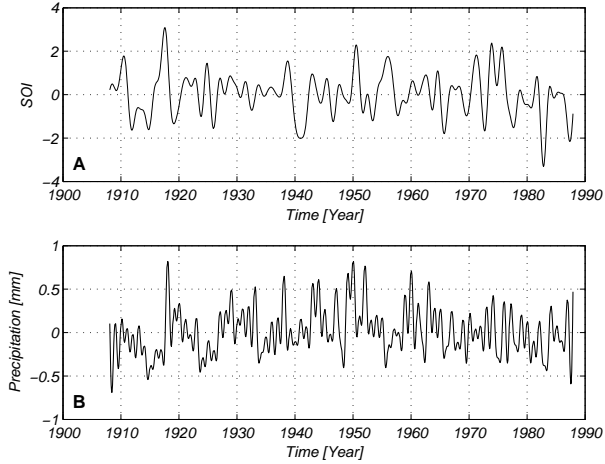, width=8cm}
\caption{(A) Southern Oscillation Index (SOI) and
(B) rainfall data of San Salvador de Jujuy.}\label{DATAjuy}
\end{figure}

\begin{figure}[tbph]
\centering \epsfig{file=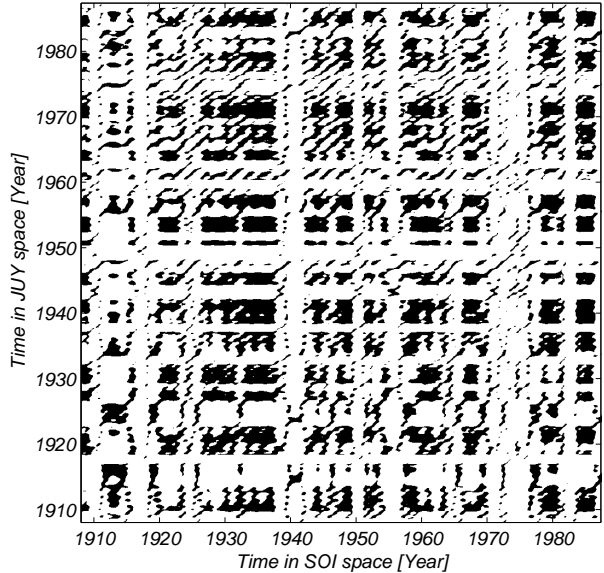,  width=8cm}
\caption{Cross recurrence plot of SOI vs.~precipitation data from the city
of San Salvador de Jujuy for an embedding of $m=3$, $\tau=4$
and $\varepsilon=1.3$).
The $x$-axis shows the time along the phase space trajectory of the SOI
and the $y$-axis that of JUY.}\label{CRPjuy}
\end{figure}

For our analysis we use monthly precipitation data
from the city San Salvador de Jujuy in NW Argentina for
the time span 1908--1987 (data from \cite{bianchi92}). The 
behaviour of the ENSO phenomenon is well represented
by the Southern Oscillation Index (SOI), which is
a normalized air pressure difference between Tahiti 
and Darwin (Fig.~\ref{DATAjuy}; data from the Climate Server of NOAA, 
1999, http://ferret.wrc.noaa.gov). Negative
extrema in SOI data mark El Ni\~no events and positive
extrema La Ni\~na events. We use the monthly SOI data for
the same time span as the rainfall data. Both data sets have
lengths of 960 points.

\begin{figure}[tbph]
\centering \epsfig{file=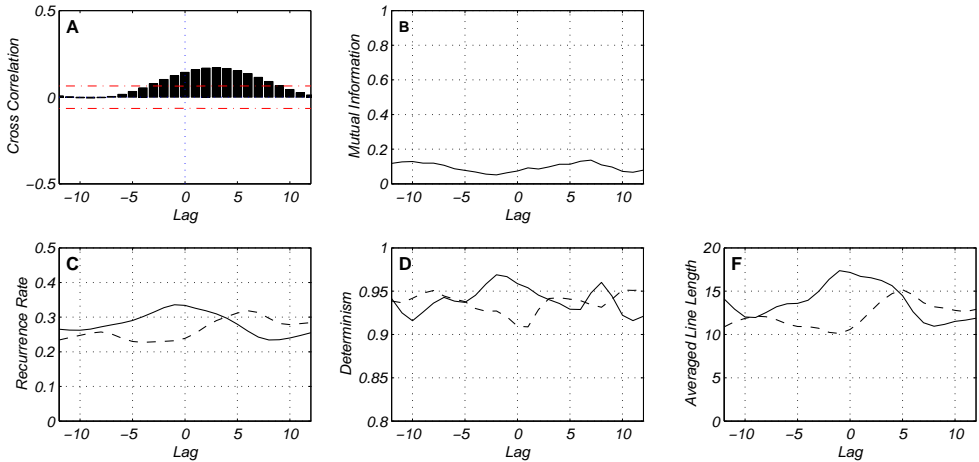,  width=13cm}
\caption{Cross correlation (A), mutual information (B) 
and CRP parameters (C, D, E) of SOI vs.~precipitation 
data from the city of San Salvador de Jujuy (JUY). 
In C, D, E, the solid lines show positive relation, the dashed lines show 
negative relation. The dashed-dotted lines in (A) mark
the 5\% confidence interval. The maxima 
of the measures reveal an interrelation between the rainfall and the
ENSO.}\label{CRQAjuy}
\end{figure}

The cross correlation function and the mutual information
show rather small correlation
$\varrho=0.14$ between both data series with time
delays of around 3 and 7 months, respectively (Fig.~\ref{CRQAjuy}A, B).

After normalization of the data, the CRP with $m=3$, 
$\tau=4$ and $\varepsilon=1.3$ is calculated and shows several structures
(Fig.~\ref{CRPjuy}).
 
The CRP analysis of local rainfall and SOI is done with a
predefined shortest diagonal length $l_{min}=6$. The analysis
reveals maxima for the complexity measures $RR_+$, $DET_+$ and
$L_+$ for correlated behaviour 
around a delay of zero months, whereas the 
measures for anti-correlated behaviour $RR_-$, $DET_-$ and
$L_-$ increase after about five months (Fig.~\ref{CRQAjuy}).
This result enables to conclude a positive relation
between ENSO and the local rainfall. This gives some
indication that the occurrence of an El Ni\~no (extreme negative SOI)
at the end of a year causes a decreased rainfall in the rainy season from
November to January and the occurrence of a La Ni\~na (extreme positive
SOI) causes an increased rainfall during this time of the year. This 
conclusion extends the results obtained by power spectra analysis, 
where the similar periodicities in both SOI and local rainfall data
were found \cite{trauth2000}. These analysis show that a
source for inter-annual precipitation variability in NW Argentina
is the ENSO \cite{trauth2000}.

The linear correlation analysis finds the correlation, however,
it is scarce above the significance and its mean at a lag of three
months. The mutual information does not reveal a clear sign
for interrelation between the data. It has small maxima at
delays of $7$ and $-10$ months.
In contrast, all the complexity measures $RR$, $DET$ and $L$
show a significant result and decomposite the correlation 
in a positive one with no delay and in a 
negative one with a delay of about five months, what suggests a more
complex interrelation between the ENSO phenomenon and local
rainfall in NW Argentina.

\section{Conclusions}

We have modified the method of cross recurrence plots (CRPs) 
in order to study the similarity of two different
phase space trajectories. Local similar time evolution of the states 
becomes then visible by long diagonal lines. The distributions of 
recurrence points and diagonal lines along the main diagonal provides an
evaluation of the similarity of the phase space trajectories of both systems.
We have introduced three measures of complexity based on
these distributions. They enable to quantify a possible similarity 
and interrelation between both dynamical systems. We have demonstrated
the potentials of this approach for typical model systems and natural 
data. In the case of linear systems, the results with this nonlinear 
technique agree with the linear correlation
test. However, in the case of nonlinear coupled systems,
the linear correlation test does not find any
correlation, whereas nonlinear techniques, as the mutual information,
and the proposed complexity measures clearly reveal this relation. Additionally,
the latters determine the kind of coupling as to be an even function. 
The application to climatological data enables to find 
a more complex relationship
between the El Ni\~no and local rainfall in NW Argentina than the
linear correlation test, the mutual information or 
the power spectra analysis yielded.

Our quantification analysis of CRPs is able to
find nonlinear relations between dynamical systems. It provides
more information than a linear correlation analysis and
the nonlinear technique of mutual information analysis. The future
work is dedicated to the development of a significance test
for RPs and the complexity measures which are based on RPs.

\section{Acknowledgments}
This work is part of the Special Research Programme {\it
Geomagnetic variations: Spatio-temporal structures, processes 
and impacts on the system Earth} and the Collaborative Research Center 
{\it Deformation Processes in the Andes} supported by the
German Research Foundation. We gratefully acknowledge 
M.\,H.~Trauth and U.~Schwarz 
for useful conversations and discussions and U.~Bahr for support of this 
work. Further we would like to thank the NOAA-CIRES
Climate Diagnostics Center for providing COADS data.

\clearpage

\bibliographystyle{elsart-num}
\bibliography{../mybibs,crp}

\end{document}